\documentclass[reprint,amssymb, amsmath, aps, superscriptaddress, showpacs, prb]{revtex4-1}
\usepackage{graphicx}
\usepackage{hyperref}
\begin{document}

\title{Strong electron-phonon coupling in the intermetallic superconductor Mo$_{8}$Ga$_{41}$}
\author{Valeriy Yu. \surname{Verchenko}}
\email{verchenko@inorg.chem.msu.ru}
\affiliation{Department of Chemistry, Lomonosov Moscow State University, 119991 Moscow, Russia}
\affiliation{National Institute of Chemical Physics and Biophysics, 12618 Tallinn, Estonia}

\author{Alexander A. \surname{Tsirlin}}
\email{altsirlin@gmail.com}
\affiliation{National Institute of Chemical Physics and Biophysics, 12618 Tallinn, Estonia}
\affiliation{Experimental Physics VI, Center for Electronic Correlations and Magnetism, Institute of Physics, University of Augsburg, 86135 Augsburg, Germany}

\author{Alexander O. \surname{Zubtsovskiy}}
\affiliation{Department of Chemistry, Lomonosov Moscow State University, 119991 Moscow, Russia}

\author{Andrei V. \surname{Shevelkov}}
\affiliation{Department of Chemistry, Lomonosov Moscow State University, 119991 Moscow, Russia}

\begin{abstract}
Crystals of superconducting Mo$_{8}$Ga$_{41}$ ($T_{\text{c}}=9.8$\,K) and Mo$_{\text{8-x}}$V$_{\text{x}}$Ga$_{41}$ limited solid solution $(x_{\text{max}}=1.9(2))$ have been grown from the high-temperature Ga flux and confirmed to have the V$_{8}$Ga$_{41}$ type of crystal structure at room temperature. Thermodynamic and transport measurements as well as electronic structure calculations were performed to investigate both normal- and superconducting-state properties of Mo$_{8}$Ga$_{41}$. The discontinuity in the heat capacity at $T_{\text{c}}$, $\Delta{}c_{\text{p}}/\gamma_{\text{N}}T_{\text{c}}=2.83$ in zero magnetic field, indicates a much stronger electron-phonon coupling than in the standard BCS limit. From the heat capacity data, we estimated the electron-phonon coupling constant $\lambda_{\text{ep}}=0.9$. The upper critical field is $\mu_{0}H_{\text{c2}}(0)=8.3$\,T, while the lower one is only $\mu_{0}H_{\text{c1}}(0)=131$\,Oe. The upper critical field of Mo$_{8}$Ga$_{41}$ exhibits a clear enhancement with respect to the Werthamer-Helfand-Honenberg prediction, consistent with the strong electron-phonon coupling. The Mo$_{\text{8-x}}$V$_{\text{x}}$Ga$_{41}$ limited solid solution also exhibits superconducting properties, and the critical temperature $T_{\text{c}}$ is reduced only slightly with increasing $x$.
\end{abstract}


\maketitle

\section{Introduction}
In previous decades, the investigation of superconductors was aimed at either finding materials with higher critical temperature $T_{\text{c}}$, or discovering non-typical superconducting behavior that significantly deviates from the conventional BCS model. Mo-based compounds attract interest in the light of both these goals. Several Mo-based carbides were reported to be superconductors with relatively high transition temperatures: $\gamma$-MoC with $T_{\text{c}}=9.3$\,K\cite{HSC}, $\eta$-MoC$_{\text{1-x}}$ (8.9\,K\cite{HSC}), $\alpha$-Mo$_2$C (6.1\,K\cite{MC}), $\beta$-Mo$_2$C (5.2\,K\cite{MC}), and Mo$_3$Al$_2$C (9.05\,K\cite{b1}). The latter compound is a noncentrosymmetric superconductor. Initially, it was ascribed to a strong-coupling regime with noticeable deviations from the BCS model\cite{b1,k1}. However, magnetic penetration depth measurements evidenced nodeless energy gap and established the conventional behavior of Mo$_3$Al$_2$C\cite{b3}. In a recent study\cite{b2}, the muon spin relaxation rate measurements did not reveal any indications of time-reversal symmetry breaking in Mo$_3$Al$_2$C and confirmed that this compound features a single \textit{s}-wave superconducting gap.

Among Mo-based compounds, Mo$_3$Sb$_7$ also attracts interest because of its complex behavior at low temperatures. Two transitions that are observed in this compound at 2.3\,K and 50\,K can be attributed to the formation of a superconducting state and spin gap, respectively\cite{m1,m2}. Therefore, Mo$_3$Sb$_7$ can be classified as a system, where superconductivity and spin fluctuations coexist, thus resembling the behavior of superconductors with strong electronic correlations\cite{m2}. The investigation of Mo$_3$Sb$_7$ by means of muon spin relaxation measurements revealed two possibilities to explain the superconducting properties: a single \textit{s}-wave superconducting gap scenario\cite{m4,m5} and a double-gap \textit{s}-wave model\cite{m6,m7}. The ambiguity between these two scenarios is so far unresolved, although the gap anisotropy and the presence of nodes are clearly excluded by the recent study of heat transport in Mo$_3$Sb$_7$ at low temperatures\cite{m8}.

Searching for other Mo-based superconductors, which may demonstrate non-typical behavior, we focused on Mo$_{8}$Ga$_{41}$ intermetallic compound. Earlier, Bezinge and Yvon\cite{m3} reported the superconductivity of Mo$_{8}$Ga$_{41}$ at $T_{\text{c}}=9.7$\,K with the upper critical field $\mu_{0}H_{\text{c2}}=8.6$\,T at zero temperature. Mo$_{8}$Ga$_{41}$ crystallizes in the centrosymmetric three-dimensional V$_{8}$Ga$_{41}$ type of crystal structure\cite{m0,m3}. The relatively high values of the transition temperature $T_{\text{c}}$ and upper critical field $\mu_{0}H_{\text{c2}}$ call for a detailed study of both normal- and superconducting-state properties that have not been addressed in the previous literature.

Here, we report crystal growth of the intermetallic Mo$_{8}$Ga$_{41}$ as well as the Mo$_{\text{8-x}}$V$_{\text{x}}$Ga$_{41}$ limited solid solution on its base. Electrical resistivity, isothermal magnetization, magnetic susceptibility and heat capacity measurements, and electronic structure calculations were performed to investigate both normal- and superconducting-state properties. By revisiting Mo$_{8}$Ga$_{41}$, we aim to provide insight into microscopic features and electronic structure of this compound.

\section{Experimental details}

\begin{figure}
\includegraphics{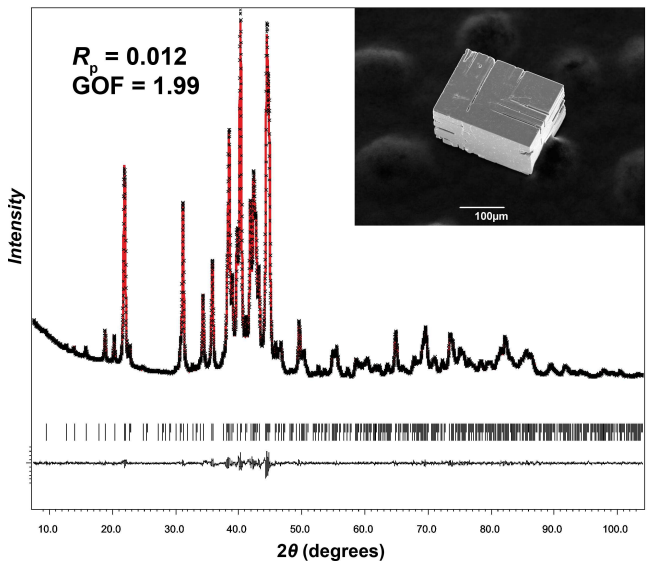}
\caption{\label{f1}Experimental (black points) and calculated (red line) XRD patterns of Mo$_{8}$Ga$_{41}$. Peak positions are given by black ticks; the difference plot is shown as a black line in the bottom part. The inset shows scanning electron microscope image of a typical Mo$_{8}$Ga$_{41}$ crystal, grown from Ga flux.}
\end{figure}

Crystals of Mo$_{\text{8-x}}$V$_{\text{x}}$Ga$_{41}$ ($x=0$, 1 and 2) were grown with a high-temperature solution growth method using Ga both as a reagent and flux medium. Mo powder (99.99\%, Sigma Aldrich), V pieces (99.9\%, Sigma Aldrich) and Ga pieces (99.999\%, Sigma Aldrich) were used as starting materials. According to the standard ampoule technique, they were weighed in the Mo:V:Ga $=(8-x):x:400$ molar ratio and placed inside quartz ampoules, which were then sealed under vacuum at pressure less than $10^{-2}$\,Torr. Ampoules were annealed in a programmable furnace at 830\,$^{\circ}$C for 55\,h and slowly cooled to 170\,$^{\circ}$C at the rate of 4\,$^{\circ}$C/h. After synthesis, the excess of liquid Ga was separated using an Eppendorf~5804R centrifuge, yielding well-shaped silvery-gray crystals (inset of Fig.~\ref{f1}). The obtained crystals were purged from the remainder of Ga metal with diluted 0.5\,M HCl during 24\,h and washed with distilled water and acetone.

Powder X-ray diffraction data were collected at room temperature using a PANalytical~X'Pert$^3$~Powder diffractometer with Cu \textit{K}$\alpha$ radiation in the $2\theta$ range between 7.5\,$^{\circ}$ and 104\,$^{\circ}$. The data were analyzed by the Rietveld method in JANA2006 software\cite{jana}.

Crystals were analyzed using a scanning electron microscope JSM~JEOL~6490-LV operated at 30\,kV and equipped with an EDX detection system INCA~x-Sight. Before operation, the EDX detection system was optimized using elemental Co as a standard. To perform quantitative elemental analysis, the system was calibrated with the use of elemental Mo and V, and GaP polished samples. All standards are provided by MAC~Analytical~Standards.

Several rectangular-shaped single crystals of Mo$_{\text{8-x}}$V$_{\text{x}}$Ga$_{41}$ were picked from the specimens with $x=0$, 1 and 2 for the subsequent crystal structure determination and refinement. Data sets were collected from the qualitatively best single crystals on a STOE STADIVARI Pilatus diffractometer equipped with a graphite monochromator and a Mo X-ray source ($\lambda_{\text{Mo}}=0.71073$\,\r A). Details of the single-crystal XRD experiments are presented in Table~S1\cite{SM}. After the numerical absorption correction, which was performed by a multiscan routine, the crystal structure was solved by direct methods\cite{sir} and refined against $F^2$ with JANA2006 software\cite{jana}. The obtained atomic coordinates and selected interatomic distances are listed in Tables~S2 and S3\cite{SM}. The atomic coordinates were standardized by the STRUCTURE~TIDY program\cite{tidy}.

Unfortunately, individual single crystals were not large enough for thermodynamic measurements. Therefore, several Mo$_{\text{8-x}}$V$_{\text{x}}$Ga$_{41}$ ($x=0$, 1 and 2) crystals were glued together and measured as polycrystalline samples. The temperature dependences of the magnetic susceptibility and isothermal magnetization were measured using a SQUID magnetometer (MPMS, Quantum Design) in magnetic fields between 0\,T and 7\,T at temperatures between 1.8\,K and 300\,K. Additionally, isothermal magnetization curves were obtained using the VSM setup of Physical Property Measurement System (PPMS, Quantum Design) in magnetic fields up to 14\,T. The heat capacity measurements were performed with a relaxation-type calorimeter (HC option, PPMS, Quantum Design) in magnetic fields between 0\,T and 11\,T at temperatures between 1.8\,K and 20\,K.

To investigate transport properties, crystals were crushed in an agate mortar and pressed into rectangular-shaped pellets at the external pressure of 100\,bars at room temperature. Densities of the pellets were estimated from their masses and linear sizes to be 87\%, 84\% and 86\% from the theoretical densities for $x=0$, 1 and 2, respectively. Electrical contacts (Cu wire, 50\,$\mu$m) were fixed using silver-containing epoxy resin (Epotek~H20E) hardened at 120\,$^{\circ}$C, and the resistance was measured by the standard four-probe method using the AC transport setup of PPMS in magnetic fields between 0\,T and 11\,T at temperatures between 1.8\,K and 400\,K with a field applied perpendicular to the direction of current.

The electronic structure of Mo$_{8}$Ga$_{41}$ was calculated within the local density approximation (LDA)\cite{lda} of the density functional theory (DFT) as implemented in the full-potential FPLO code (version 14.00-47)\cite{fplo} with the basis set of local orbitals. The integrations in \textit{k}-space were performed by the improved tetrahedron method\cite{tetr} on a grid of $16\times16\times16$ \textit{k}-points for the scalar relativistic calculation. The density of states at the Fermi level in the case of Mo$_{\text{8-x}}$V$_{\text{x}}$Ga$_{41}$ solid solution was calculated using the rigid band shift approximation.

\section{RESULTS AND DISCUSSION}
\subsection{Crystal structure of Mo$_{8}$Ga$_{41}$}

\begin{figure}
\includegraphics{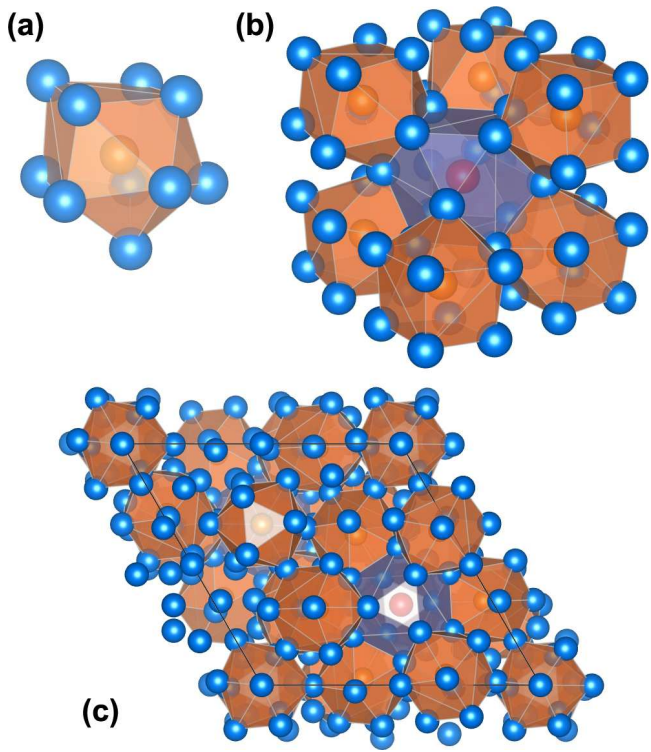}
\caption{\label{f2}Mo$_{8}$Ga$_{41}$ crystal structure: (a) MoGa$_{10}$ polyhedron (orange), (b) the main building block, constructed by MoGa$_{10}$ polyhedra (orange) and GaGa$_{12}$ cuboctahedra (blue), (c) the unit cell content.}
\end{figure}

The powder XRD pattern of crushed crystals of Mo$_{8}$Ga$_{41}$ is shown in Fig.~\ref{f1}. All lines in the pattern could be indexed in the $R\bar{3}$ (\# 148) space group, confirming  phase purity of the sample. The respective Rietveld refinement of the data yields the cell parameters $a=14.0454(2)$\,\r A and $c=15.0525(2)$\,\r A, which are in good agreement with the previously reported values\cite{m0}.

A suitable single crystal with the composition Mo$_{8.1(2)}$Ga$_{40.9(2)}$, confirmed by EDX spectroscopy, was selected from the reaction products to perform structure determination. The collected data could be indexed with the cell parameters $a=14.0290(9)$\,\r A and $c=15.0414(9)$\,\r A in the $R\bar{3}$ (\# 148) space group, which was chosen on the basis of systematic extinction conditions. Atomic positions were determined by direct methods and refined in full-matrix anisotropic approximation. Details of the data collection and refinement are summarized in Table~S1, the atomic parameters and selected interatomic distances are given in Tables~S2 and S3, respectively. Our refinement shows that Mo$_{8}$Ga$_{41}$ crystallizes in the V$_{8}$Ga$_{41}$ type of crystal structure, in agreement with the previously reported results\cite{m0,m3}. The refinement indicates the absence of partially occupied positions in the crystal structure, thus confirming that Mo$_{8}$Ga$_{41}$ is a stoichiometric compound. 

In the Mo$_{8}$Ga$_{41}$ crystal structure, Mo and Ga atoms occupy two and nine crystallographic positions, respectively. Mo atoms are well separated from each other, and, consequently, only Ga atoms appear in the first coordination sphere of Mo, which is presented by MoGa$_{10}$ polyhedra for both the Mo1 and Mo2 positions.  The MoGa$_{10}$ polyhedron (Fig.~\ref{f2}a) consists of one half of a MoGa$_{8/2}$ cube and one half of a MoGa$_{12/2}$ icosahedron. MoGa$_{10}$ polyhedra are interconnected by corners and form the arrangement, in which one triangular face of each polyhedron is shared with a cuboctahedron centered by the unique Ga atom (Fig.~\ref{f2}b). The Ga atom in the centre of a cuboctahedron (crystallographic position 3\textit{b} (0; 0; 0.5)) has no contacts with Mo atoms in the first coordination sphere. Eight MoGa$_{10}$ polyhedra are condensed on the faces of a cuboctahedron, which is occupied by one Ga atom, yielding the $8\text{MoGa}_{10/2}+1\text{Ga}=\text{Mo}_8\text{Ga}_{41}$ composition of the compound. This arrangement represents a building block of the Mo$_{8}$Ga$_{41}$ crystal structure, which is clearly three-dimensional with Mo atoms being evenly distributed in the matrix of Ga atoms (Fig.~\ref{f2}c).

\subsection{Electrical resistivity}

\begin{figure}
\includegraphics{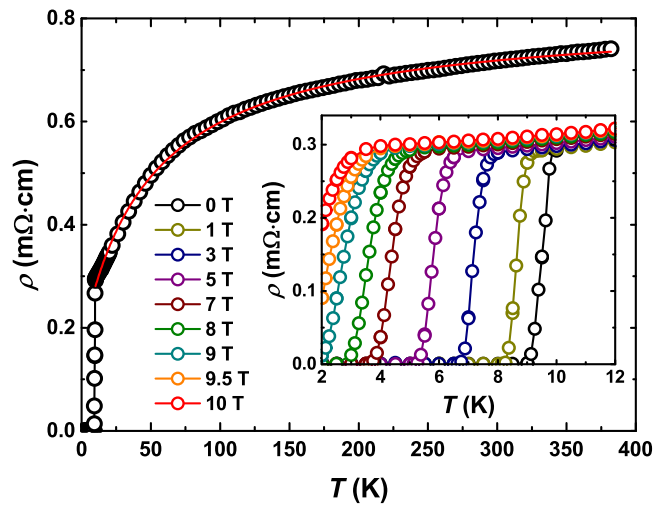}
\caption{\label{f3}Electrical resistivity of Mo$_{8}$Ga$_{41}$ in zero magnetic field. The solid curve represents a fit according to the parallel resistance model. The inset shows resistivity at low temperatures in different applied fields.}
\end{figure}

The temperature dependence of the Mo$_{8}$Ga$_{41}$ electrical resistivity, measured in zero  magnetic field, is shown in Fig.~\ref{f3}. The $\rho(T)$ exhibits metallic behavior with rather high absolute values of resistivity well above the Mott-Ioffe-Regel limit ($\sim0.1$\,m$\Omega$\,cm). Moreover, at temperatures above 75\,K the resistivity grows much slower than at low temperatures. Such behavior was observed in many metallic systems with large resistivity\cite{r1}, particularly, in strong-coupled transition-metal A15 superconductors\cite{r2}. The saturation of resistivity in these compounds occurs when the mean electron free path is comparable to the interatomic distances\cite{r1,r2}. To account for the saturation of resistivity at high temperatures, the parallel resistance model was used, in which $\frac{1}{\rho(T)}=\frac{1}{\rho_{\text{sat}}}+\frac{1}{\rho_{\text{ideal}}(T)}$. We used a linear behavior for the ideal resistivity: $\rho_{\text{ideal}}(T)=\rho_{0}+\alpha{}T$. A fit employing this model is shown in Fig.~\ref{f3} by the solid red line and yields $\rho_{0}=0.221(5)$\,m$\Omega$\,cm and $\rho_{\text{sat}}=0.808(2)$\,m$\Omega$\,cm.

At low temperatures, the resistivity of Mo$_{8}$Ga$_{41}$ clearly indicates a superconducting transition at $T_{\text{c}}=9.8$\,K, where $T_{\text{c}}$ was determined as an onset temperature of the sharp resisitivity drop. The obtained value of $T_{\text{c}}$ is in agreement with the data reported by Bezinge and Yvon\cite{m3}. The inset of Fig.~\ref{f3} shows the $\rho(T)$ around the transition in various magnetic fields. As expected, magnetic field shifts the superconducting transition to lower temperatures. These data were used to extract the upper critical field at the transition temperature $T_{\text{c}}$, which was determined as an onset temperature of the resistive drop. The upper critical field $\mu_{0}H_{\text{c2}}(T)$ of Mo$_{8}$Ga$_{41}$ is shown in Fig.~\ref{f8} and will be discussed further in the text.

\subsection{Magnetic susceptibility}

\begin{figure}
\includegraphics{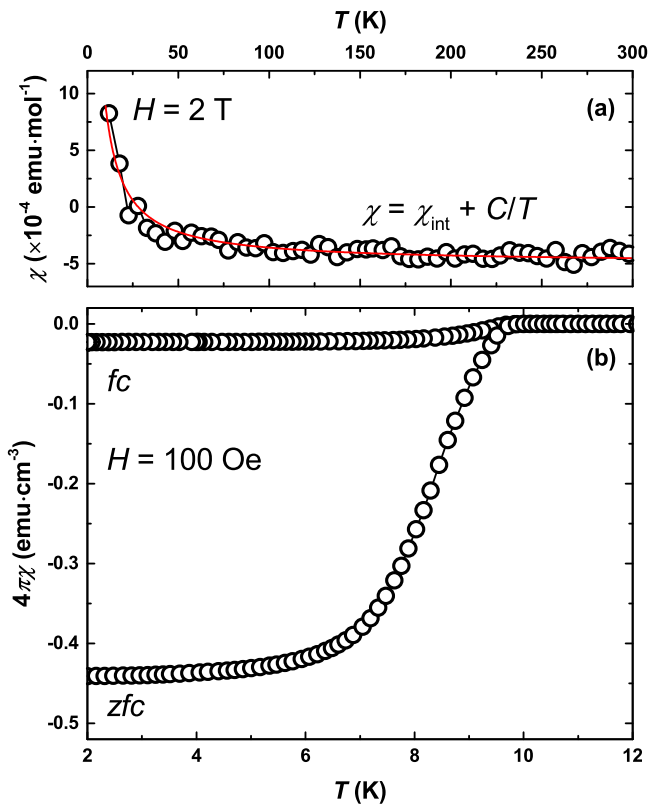}
\caption{\label{f4}(a) Normal-state magnetic susceptibility of Mo$_{8}$Ga$_{41}$ measured at $H=2$\,T; (b) magnetic susceptibility at $H=100$\,Oe, measured in zfc and fc conditions.}
\end{figure}

The normal-state magnetic susceptibility of Mo$_{8}$Ga$_{41}$, measured in the temperature range between 10\,K and 300\,K in the $H=2$\,T magnetic field, is shown in Fig.~\ref{f4}a. The compound exhibits bulk diamagnetism with the upturn of the magnetic sucseptibility at low temperatures $T<100$\,K, which is most likely due to paramagnetic impurities. Temperature dependence of the magnetic susceptibility can be satisfactorily fitted with the equation $\chi(T)=\chi_{\text{int}}+C/T$, resulting in the intrinsic susceptibility $\chi_{\text{int}}=-4.96(8)\times10^{-4}$\,emu\,$\text{mol}^{-1}$ and $C=0.0139(4)$\,emu\,$\text{K}$\,$\text{mol}^{-1}$. The intrinsic susceptibility $\chi_{\text{int}}$ includes the following summands: $\chi_{\text{int}}=\chi_{\text{P}}+\chi_{\text{core}}+\chi_{\text{VV}}+\chi_{\text{L}}$, where $\chi_{\text{P}}$ is the Pauli paramagnetic spin susceptibility of the conduction electrons, $\chi_{\text{core}}$ is the diamagnetic orbital contribution from the electrons (ionic or atomic), $\chi_{\text{VV}}$ is the Van Vleck paramagnetic orbital contribution, and $\chi_{\text{L}}$ is the Landau orbital diamagnetism of the conduction electrons. A realistic estimation of $\chi_{\text{P}}$ from the above equation is problematic, mostly because of uncertainty in the calculation of $\chi_{\text{core}}$ value. For instance, the $\chi_{\text{core}}$ value for Mo ions gradually decreases from $-7\times10^{-6}$\,emu\,$\text{mol}^{-1}$ for $\text{Mo}^{6+}$ to $-31\times10^{-6}$\,emu\,$\text{mol}^{-1}$ for $\text{Mo}^{2+}$ species\cite{dia}. However, chemical bonding in Mo$_{8}$Ga$_{41}$ is neither ionic nor covalent, therefore, the core contribution is hard to estimate precisely. 

The sample contains minor amount of paramagnetic impurities resulting in a Curie-like $C/T$ contribution to $\chi$. The obtained value of $C$ is equal to the presence of 3.7\,mol.\% of the $S=\frac12$ paramagnetic impurity assuming $g=2$. This impurity contribution can be attributed to the defects induced by the post-synthetic treatment of Mo$_{8}$Ga$_{41}$ crystals with diluted 0.5M HCl. However, these impurities should not affect other physical properties, such as electrical resistivity and heat capacity of Mo$_{8}$Ga$_{41}$.

The temperature dependence of the zero-field-cooled (zfc) and field-cooled (fc) magnetic susceptibilities of Mo$_{8}$Ga$_{41}$, measured in 100\,Oe magnetic field, is shown in Fig.~\ref{f4}b. The divergence of the zfc and fc curves indicates a transition to the superconducting state at $T_{\text{c}}=9.8$\,K, in good agreement with the resistivity data. The fc signal (Meissner effect) is weak, which is most probably due to strong flux line pinning in this type-II superconductor. For the zfc signal, the transition is significantly broadened with temperature, having the width of about 5\,K. The reduced volume susceptibility $4\pi\chi$ reaches $-0.44$\,emu$\,\text{cm}^{-3}$ at $T=2$\,K, however, the data were not corrected for the demagnetization effect, since the sample shape is not well-defined. Assuming the complete diamagnetic response of the specimen, one obtains the demagnetization factor of 2.3, which is somewhat larger than the value of 1.0 expected for an ellipsoid of revolution. 

The isothermal magnetization data were used to estimate both lower $\mu_{0}H_{\text{c1}}(T)$ and upper $\mu_{0}H_{\text{c2}}(T)$ critical fields of Mo$_{8}$Ga$_{41}$ (Fig.~\ref{f8}). $\mu_{0}H_{\text{c1}}(T)$ was determined as a field, at which the $M(H)$ dependence deviates from a linear behavior with the probability of 95\% in the low-field $M(H)$ data. $\mu_{0}H_{\text{c2}}(T)$ was determined as a kink on the high-field $M(H)$ curve.

\subsection{Heat capacity}

\begin{figure}
\includegraphics{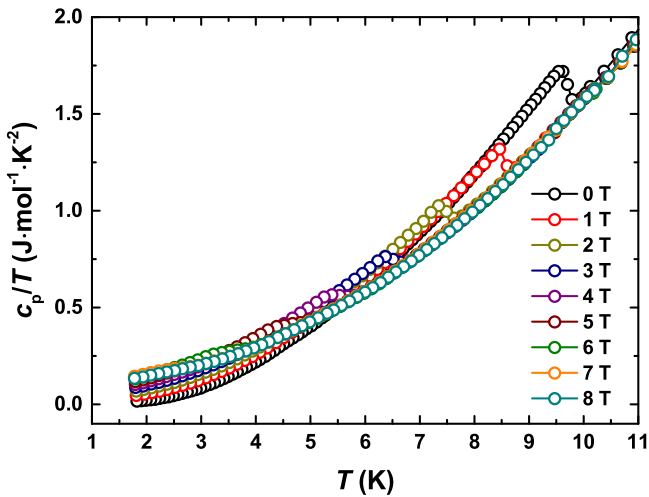}
\caption{\label{f5}Specific heat $c_{\text{p}}/T$ versus $T$ plot in various magnetic fields for Mo$_{8}$Ga$_{41}$.}
\end{figure}

The specific heat $c_{\text{p}}/T$ versus $T$ plot in the temperature range between 1.8\,K and 11\,K in various magnetic fields is shown in Fig.~\ref{f5}. A sharp anomaly with the transition temperature $T_{\text{c}}=9.7$\,K is observed in zero magnetic field, confirming bulk superconductivity of Mo$_{8}$Ga$_{41}$. The temperature of the transition was determined by a graphical equal-areas approximation (entropy-conserving) for each field $H$. Magnetic field shifts the transition to lower temperatures, and finally, no sign of the transition is observed in the $H=10$\,T magnetic field at temperatures above 1.8\,K. Using the data obtained in different magnetic fields, we calculated the upper critical field $\mu_{0}H_{\text{c2}}(T)$ of Mo$_{8}$Ga$_{41}$ (Fig.~\ref{f8}).

\begin{figure}
\includegraphics{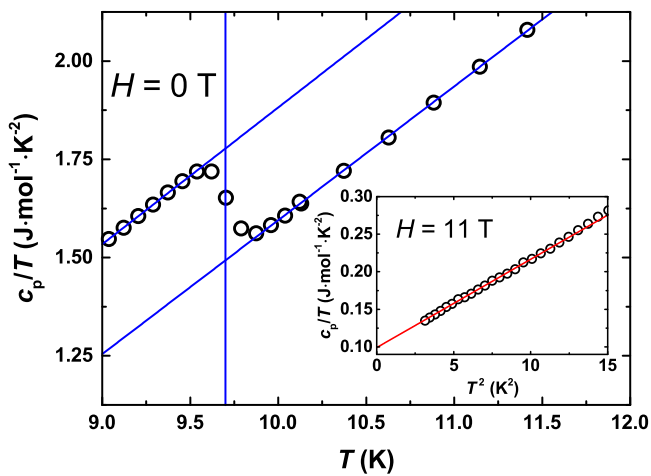}
\caption{\label{f6}The specific heat anomaly in zero magnetic field. The solid curve is a construction to estimate the specific heat jump at $T=T_{\text{c}}$. The inset shows $c_{\text{p}}/T$ versus $T^2$ plot in $H=11$\,T magnetic field, the solid curve represents the linear fit of the data.}
\end{figure}

A closer view at the specific heat anomaly in zero magnetic field is presented in Fig.~\ref{f6}. We used line approximation of the data just below and above the transition to extract the value of the jump in the specific heat at $T_{\text{c}}$. The corresponding construction is shown by blue lines in Fig.~\ref{f6}, with the vertical blue line located at the transition temperature $T_{\text{c}}=9.7$\,K. According to this approximation, we obtained $\Delta{}c_{\text{p}}/T_{\text{c}}=280$\,mJ\,$\text{mol}^{-1}$\,$\text{K}^{-2}$, although the $\Delta{}c_{\text{p}}/T_{\text{c}}$ value might be a bit overestimated by this method due to the smoothness of the superconducting transition.

The normal-state specific heat of Mo$_{8}$Ga$_{41}$, measured in the 11\,T magnetic field, is shown in the inset of Fig.~\ref{f6}. The $c_{\text{p}}/T$ versus $T^2$ plot can be fitted by the equation $c_{\text{p}}/T=\gamma_{\text{N}}+\beta{}T^{2}$ at temperatures between 1.77\,K and 3.53\,K, giving $\gamma_{\text{N}}=99.1(5)$\,mJ\,$\text{mol}^{-1}$\,$\text{K}^{-2}$ and $\beta=11.76(7)$\,mJ\,$\text{mol}^{-1}$\,$\text{K}^{-4}$, which yields the Debye temperature of $\Theta_{\text{D}}=201$\,K. Thus, the normalized jump of the specific heat at $T_{\text{c}}$ is $\Delta{}c_{\text{p}}/\gamma_{\text{N}}T_{\text{c}}=2.83$. This value indicates a much stronger electron-phonon coupling in the superconducting state of Mo$_{8}$Ga$_{41}$ than the weak-coupling BCS limit, where $\Delta{}c_{\text{p}}/\gamma_{\text{N}}T_{\text{c}}=1.43$ is expected. The electron-phonon coupling constant $\lambda_{\text{ep}}$ can be estimated using the equation:

\begin{equation}
\lambda_{\text{ep}}=\frac{\gamma_{\text{N}}}{\gamma_{\text{bare}}}-1,\label{eq1}
\end{equation}

\noindent{}where $\gamma_{\text{N}}$ and $\gamma_{\text{bare}}=\frac{\pi^{2}k_{\text{B}}^2}{3}N(E_{\text{F}})$ are the experimental and calculated values of the Sommerfield coefficient of the electronic specific heat, respectively, and $N(E_{\text{F}})$ is the density of states at the Fermi level. From the band structure calculations we obtained $N(E_{\text{F}})=22.36$\,st.\,$\text{eV}^{-1}$\,$\text{f.u.}^{-1}$, yielding $\gamma_{\text{bare}}=52.7$\,mJ\,$\text{mol}^{-1}$\,$\text{K}^{-2}$ and $\lambda_{\text{ep}}=0.88$. In addition, $\lambda_{\text{ep}}$ can be estimated in the single-gap superconductivity approximation using McMillan's formula\cite{MM}:

\begin{equation}
\lambda_{\text{ep}}=\frac{1.04+\mu^{*}\ln\left(\frac{\Theta_{\text{D}}}{1.45T_{\text{c}}}\right)}{(1-0.62\mu^{*})\ln\left(\frac{\Theta_{\text{D}}}{1.45T_{\text{c}}}\right)-1.04},\label{eq2}
\end{equation}

\begin{figure}
\includegraphics{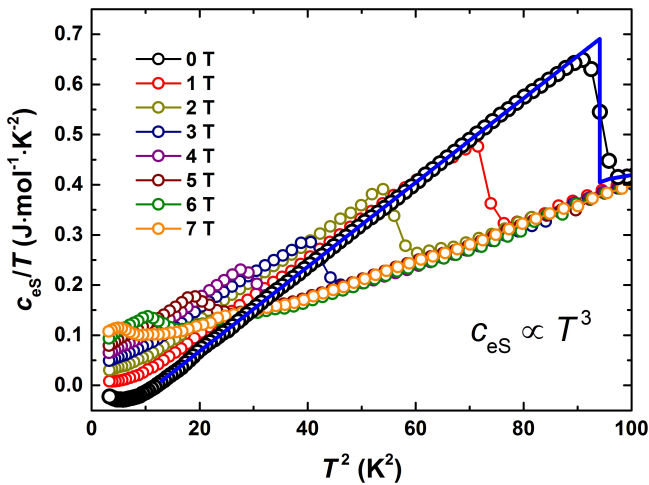}
\caption{\label{f7}Electronic specific heat $c_{\text{eS}}/T$ versus $T^2$ plot in various magnetic fields together with $\propto{}T^3$ behavior for $H=0$\,T, which is shown by the solid blue line.}
\end{figure}

\noindent{}where $\mu^{*}$ is the Coulomb pseudopotential, $\Theta_{\text{D}}$ is the Debye temperature, and $T_{\text{c}}$ is the superconducting transition temperature. The Coulomb pseudopotential, $\mu^{*}$, is equal to 0.1 in the case of nearly-free-electron metals, as confirmed empirically for Zn. For many other transition metals, the empirical values of $\mu^{*}$ vary only slightly around 0.13, therefore, $\mu^{*}=0.1$ and 0.15 values could be used for estimating the plausible range of $\lambda_{\text{ep}}$\cite{MM}. The values of $\Theta_{\text{D}}=201$\,K and $T_{\text{c}}=9.7$\,K, obtained above from the heat capacity data, yield $\lambda_{\text{ep}}=0.90$ and 1.06 for $\mu^{*}=0.1$ and 0.15, respectively. Thus, we find a good agreement between the $\lambda_{\text{ep}}$ values, calculated using equations (\ref{eq1}) and (\ref{eq2}). Henceforth, we will take $\lambda_{\text{ep}}=0.9$, which corresponds to $\mu^{*}=0.1$ and $\gamma_{\text{bare}}=52.7$\,mJ\,$\text{mol}^{-1}$\,$\text{K}^{-2}$.

The obtained high value of $\lambda_{\text{ep}}$ implies strong electron-phonon coupling in the superconducting state of Mo$_{8}$Ga$_{41}$. Together with the unusually high value of $\Delta{}c_{\text{p}}/\gamma_{\text{N}}T_{\text{c}}=2.83$, it indicates that the superconductivity of Mo$_{8}$Ga$_{41}$ significantly deviates from the BCS limit. To analyze the superconducting state in more detail, we calculated the electronic contribution to the heat capacity by subtracting the $\beta{}T^3$ lattice term from the $c_{\text{p}}(T)$ data. The resulting $c_{\text{eS}}/T$ versus $T^2$ plot in various magnetic fields is shown in Fig.~\ref{f7}. Unfortunately, the subtraction resulted in negative values of $c_{\text{eS}}$ in zero magnetic field below 3.5\,K, which is, without doubt, due to the loss of accuracy in the estimation of $\beta$. However, negative values of $c_{\text{eS}}$ are not observed in higher magnetic fields. Moreover, we found that $c_{\text{eS}}(T)\propto{}T^3$ in the superconducting state at temperatures above 3\,K (for $H=0$\,T the corresponding construction is shown by blue line in Fig.~\ref{f7}). This additionally evidences non-BCS-type superconductivity of Mo$_{8}$Ga$_{41}$. At low temperatures, the heat capacity data exhibit an anomaly, which is seen as the upturn of $c_{\text{eS}}$ in zero magnetic field. This anomaly may be attributed either to a small fraction of impurities in the specimen or to a non-trivial gap structure of the superconducting state. Due to the low-temperature anomaly, it remains unclear whether or not $c_{\text{eS}}$ of Mo$_{8}$Ga$_{41}$ is compatible with a nodal structure.

\subsection{Lower and upper critical magnetic fields}

\begin{figure}
\includegraphics{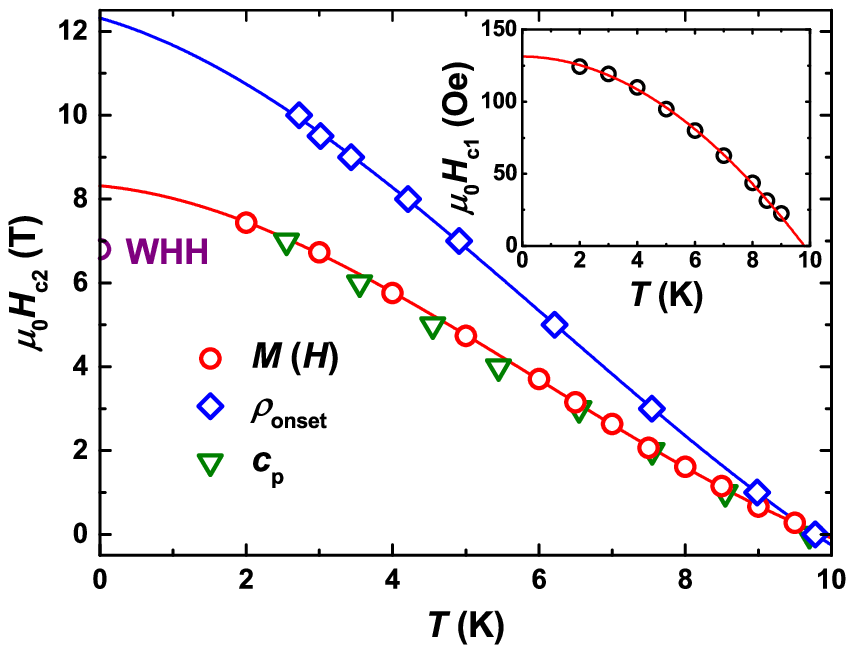}
\caption{\label{f8}Upper critical field $\mu_{0}H_{\text{c2}}(T)$ of Mo$_{8}$Ga$_{41}$, determined from the magnetization, heat capacity and electrical resistivity measurements. The solid lines represent an extrapolation of the data by the second-order polynomial. Lower critical field $\mu_{0}H_{\text{c1}}(T)$ is shown in the inset. The solid red line is a fit of the data, as described in the text.}
\end{figure}

From the magnetic field dependences of magnetization, electrical resistivity and heat capacity, we obtained the upper critical field of Mo$_{8}$Ga$_{41}$, which is shown in Fig.~\ref{f8}. The $\mu_{0}H_{\text{c2}}(T)$ data from $M(H)$ and $c_{\text{p}}$ perfectly coincide. The deviation of the upper critical field values, obtained from the resistivity measurements, is likely due to the influence of surface superconductivity, which results in $\mu_{0}H_{\text{c3}}(T)$. Regarding details of the resistivity measurements, we expect surface superconductivity emerging in each cross-sectional area of the sample, thus influencing the upper critical field substantially. From a free extrapolation by the second-order polynomial, we obtained $\mu_{0}H_{\text{c3}}(0)=12.3$\,T from $\rho_{\text{onset}}$ and $\mu_{0}H_{\text{c2}}(0)=8.3$\,T from $M(H)$, yielding $\mu_{0}H_{\text{c3}}(0)/\mu_{0}H_{\text{c2}}(0)=1.5$. In the ideal case $\mu_{0}H_{\text{c3}}(0)/\mu_{0}H_{\text{c2}}(0)=1.7$\cite{hc3}, however, deviations are known for a number of type-II superconducting alloys, for which the values of $\mu_{0}H_{\text{c3}}(0)/\mu_{0}H_{\text{c2}}(0)$ between 1.4 and 2.1 are found experimentally\cite{hc3,hc32}.

We will further discuss $\mu_{0}H_{\text{c2}}(T)$ from $M(H)$ and $c_{\text{p}}$, only. All the curves vary linearly with $T$ down to $0.4T_{\text{c}}$, thus, the upper critical field $\mu_{0}H_{\text{c2}}(0)$ can be estimated using the Werthamer-Helfand-Honenberg (WHH) formula for the clean limit: $\mu_{0}H_{\text{c2}}(0)=-0.693T_{\text{c}}\left(\left.\frac{d\mu_{0}H_{\text{c2}}(T)}{dT}\right|_{T_{\text{c}}}\right)$\cite{whh}. Using the values of $\left.\frac{d\mu_{0}H_{\text{c2}}(T)}{dT}\right|_{T_{\text{c}}}=-1$\,T/K and $T_{\text{c}}=9.8$\,K, we obtained $\mu_{0}H_{\text{c2}}(0)=6.8$\,T. Alternatively, a free extrapolation by the second-order polynomial, which is shown by the red line in Fig.~\ref{f8}, leads to the estimated value of the upper critical field $\mu_{0}H_{\text{c2}}(0)=8.3$\,T from the $M(H)$ data. The observed value shows a clear enhancement with respect to the Werthamer-Helfand-Honenberg prediction, consistent with the strong electron-phonon coupling inferred from the the jump in the specific heat. The $\mu_{0}H_{\text{c2}}(0)$ value corresponds to the Ginzburg-Landau coherence length of $\xi_{\text{GL}}=158$\,\r A, as calculated from $\mu_{0}H_{\text{c2}}(0)=\frac{\Phi_0}{2\pi\xi_{\text{GL}}^2}$, where $\Phi_0$ is the flux quantum $h/2e$. $\mu_{0}H_{\text{c2}}(0)$ is significantly lower than the Pauli-paramagnetic limit for weak electron-phonon coupling, which is $\mu_{0}H_{\text{P}}=1.86T_{\text{c}}=18.2$\,T. The obtained value of $\mu_{0}H_{\text{P}}$ yields the Maki parameter $\alpha=\sqrt{2}\mu_{0}H_{\text{c2}}(0)/\mu_{0}H_{\text{P}}=0.53$, where $\mu_{0}H_{\text{c2}}(0)$ is the WHH limit of the upper critical field. Alternatively, the Maki parameter $\alpha$ can be derived from $\gamma_{\text{N}}$ and $\rho_{0}$ using the formula applicable in the dirty limit\cite{whh}:

\begin{equation}
\alpha=\frac{3e^{2}\hbar\gamma_{\text{N}}\rho_{0}}{2m_{\text{e}}\pi^{2}k_{\text{B}}^{2}V_{\text{m}}},
\end{equation}

\noindent{}where $\gamma_{\text{N}}$ is the experimental value of the Sommerfield coefficient, $\rho_{0}$ is the normal-state resistivity extrapolated to zero temperature, and $V_{\text{m}}$ is the molar volume. Using the values of $\gamma_{\text{N}}=99.1(5)$\,mJ\,$\text{mol}^{-1}$\,$\text{K}^{-2}$ and $\rho_{0}=0.221(5)$\,m$\Omega$\,cm, we obtained $\alpha=1.02$. The difference between the two estimates of $\alpha$ may arise from uncertainties in determining the absolute value of resistivity $\rho_{0}$ on a polycrystalline sample. Nevertheless, both estimates of $\alpha$ indicate that the SC transition should be a second-order phase transition, as expected for a type-II superconductor.

The inset of Fig.~\ref{f8} shows the lower critical field $\mu_{0}H_{\text{c1}}(T)$ of Mo$_{8}$Ga$_{41}$ obtained from the magnetization data. The $\mu_{0}H_{\text{c1}}(T)$ data can be satisfactory fitted using the equation $\mu_{0}H_{\text{c1}}(T)=\mu_{0}H_{\text{c1}}(0)(1-\left(\frac{T}{T_{\text{c}}}\right)^\beta)$. The fitting curve is presented as a red line in the inset of Fig.~\ref{f8}. Assuming that the transition temperature is $T_{\text{c}}=9.8$\,K, we obtained $\mu_{0}H_{\text{c1}}(0)=131(2)$\,Oe and $\beta=1.95(5)$. The lower critical field $\mu_{0}H_{\text{c1}}(0)$, which indicates the appearance of flux lines in the sample volume, is much lower than the upper critical field $\mu_{0}H_{\text{c2}}(0)$ of Mo$_{8}$Ga$_{41}$, as it is usually observed in type-II superconductors.

\subsection{Electronic structure}

\begin{figure}
\includegraphics{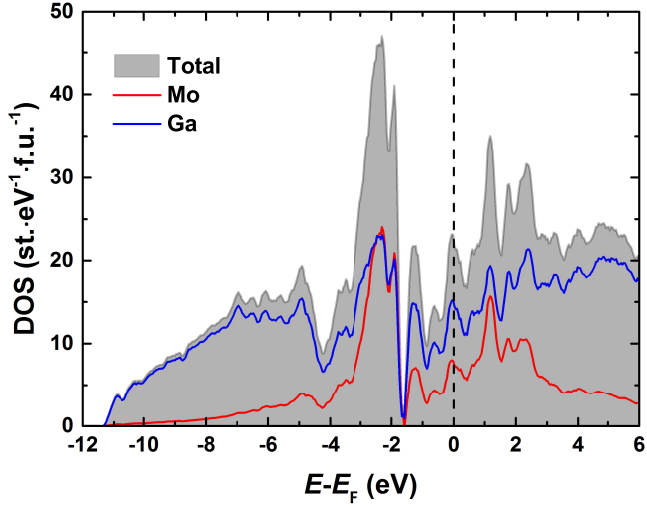}
\caption{\label{f9}Calculated electronic density of states (DOS) for Mo$_{8}$Ga$_{41}$. The Fermi level is indicated by the dashed line.}
\end{figure}

The calculated electronic density of states (DOS) for Mo$_{8}$Ga$_{41}$ is shown in Fig.~\ref{f9}. In general, the energy spectrum of Mo$_{8}$Ga$_{41}$ is similar to those calculated for the isostructural compounds V$_{8}$Ga$_{41}$ and T$_{8}$Ga$_{41-\text{y}}$Zn$_{\text{y}}$\cite{h1} (T = V, Cr, Mn). The energetically low-lying region between $-11$ and $-4$\,eV is primarily composed by Ga 4\textit{s} and 4\textit{p} contributions with a small admixture of Mo 4\textit{d} states. At higher energies above $-4$\,eV, the strong bonding between Ga and Mo species is observed. The mixing of Ga 4\textit{p} and Mo 4\textit{d} orbitals results in the sharp peak of the DOS reaching almost 50\,st.\,$\text{eV}^{-1}$\,$\text{f.u.}^{-1}$ at the relative energy of $-2.3$\,eV. This peak is separated from the rest of the spectrum by the dip, located at $-1.6$\,eV. The DOS near the Fermi level at energies between $-1.6$\,eV and 3\,eV also has a sharp peak structure with the Fermi level located close to the local maximum of the DOS, composed mainly of Ga 4\textit{p} and Mo 4\textit{d} contributions. The resulting high value of the DOS at the Fermi level is $N(E_{\text{F}})=22.36$\,st.\,$\text{eV}^{-1}$\,$\text{f.u.}^{-1}$ yields the bare Sommerfield coefficient of the electronic specific heat $\gamma_{\text{bare}}=52.7$\,mJ\,$\text{mol}^{-1}$\,$\text{K}^{-2}$. 

In the Mo$_{8-\text{x}}$V$_{\text{x}}$Ga$_{41}$ solid solution, the substitution of V for Mo results in the reduced number of valence electrons per formula unit. Consequently, the Fermi level shifts to lower energies resulting in the density of states at the Fermi level $N(E_{\text{F}})=25.96$\,st.\,$\text{eV}^{-1}$\,$\text{f.u.}^{-1}$ and 23.25\,st.\,$\text{eV}^{-1}$\,$\text{f.u.}^{-1}$ for $x=1$ and 1.9, respectively. These values only slightly exceed the $N(E_{\text{F}})$ value obtained for the unsubstituted Mo$_{8}$Ga$_{41}$. As a result, the DOS at the Fermi level remains high, thus remaining favorable for the superconductivity. Moreover, a moderate increase of $T_{\text{c}}$ is expected from the derived values of $N(E_{\text{F}})$.

\subsection{Mo$_{\text{8-x}}$V$_{\text{x}}$Ga$_{41}$ solid solution}

\begin{figure}
\includegraphics{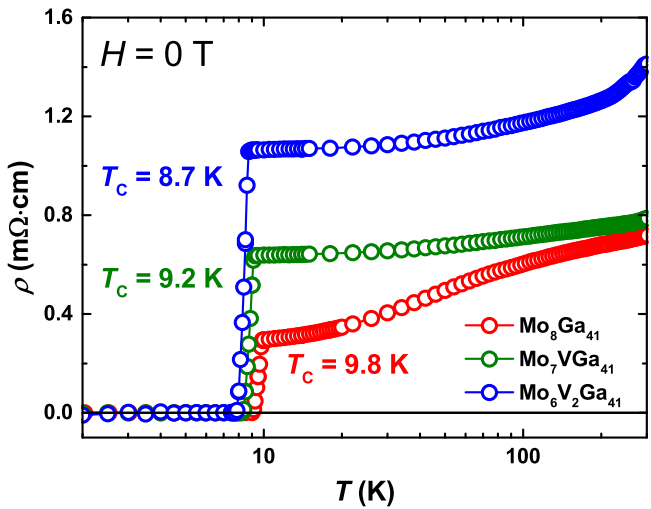}
\caption{\label{f10}Electrical resistivity of the Mo$_{\text{8-x}}$V$_{\text{x}}$Ga$_{41}$ solid solution for $x=0$, 1 and 2 in zero magnetic field.}
\end{figure}

The fact that Mo$_{8}$Ga$_{41}$ and V$_{8}$Ga$_{41}$ intermetallic compounds possess the same type of crystal structure motivated us to investigate the formation of the Mo$_{\text{8-x}}$V$_{\text{x}}$Ga$_{41}$ solid solution. Crystals of the solid solution were obtained by the high-temperature gallium flux technique using the Mo:V molar ratio of 7:1 and 6:2 in the reaction mixture for $x=1$ and 2, respectively. The obtained crystals were crushed and analyzed by the standard X-ray diffraction technique, which showed that the specimen with $x=1$ is single-phase, and its diffraction pattern is consistent with the Mo$_{8}$Ga$_{41}$ structure type. Moreover, the Mo$_{7.0(2)}$V$_{1.0(1)}$Ga$_{41.0(2)}$ elemental composition was determined by EDX spectroscopy, and it perfectly agrees with the nominal one. 

According to the XRD results, the specimen with $x=2$ also represents the Mo$_{8}$Ga$_{41}$-based solid solution, but additionally it contains a small admixture of the secondary V$_{8}$Ga$_{41}$-based phase. For this specimen, the averaged elemental composition of crystals was found to be Mo$_{6.1(1)}$V$_{1.9(2)}$Ga$_{41.0(1)}$ by EDXS. All these results suggest the formation of the limited solid solution Mo$_{\text{8-x}}$V$_{\text{x}}$Ga$_{41}$ with $x_{\text{max}}=1.9(2)$. 

Single-crystal XRD experiments were carried out to investigate the crystal structure of the solid solution at room temperature. Details of the crystal data collection and refinement are summarized in Table~S1, the obtained atomic parameters and selected interatomic distances are given in Tables~S2 and S3, respectively. According to the single-crystal XRD results, the solid solution crystallizes in the parent V$_{8}$Ga$_{41}$ type of crystal structure. V atoms substitute Mo atoms almost evenly in two crystallographic positions, as it is seen from the refined occupation factors (s.o.f. in Table~S2). Note that the total s.o.f. values were constrained to achieve the compositions determined by EDXS for each sample.

Thermodynamic and transport measurements suggest that the solid solution becomes superconducting at low temperatures, very similar to the parent Mo$_{8}$Ga$_{41}$ compound. The $\rho(T)$ curves measured in zero magnetic field for $x=0$, 1 and 2 are given in Fig.~\ref{f10}. As in the case of Mo$_{8}$Ga$_{41}$, the solid solution exhibits metallic behavior with high absolute values of resistivity. The thermodynamic and transport measurements confirm bulk superconductivity of the Mo$_{\text{8-x}}$V$_{\text{x}}$Ga$_{41}$ solid solution with $T_{\text{c}}=9.8$\,K, 9.2\,K and 8.7\,K in zero magnetic field for $x=0$, 1 and 2, respectively. Thus, the sizable substitution of V for Mo atoms in the crystal structure has rather weak effect on superconductivity, as expected from the high values of $N(E_{\text{F}})$ for the solid solution. However, the reduction in $T_{\text{c}}$ do not correlate with the evolution of $N(E_{\text{F}})$, since the disorder induced by the random arrangement of Mo and V atoms in the crystal structure may also have impact on $T_{\text{c}}$.

\section{Conclusions}
Crystals of the intermetallic compound Mo$_{8}$Ga$_{41}$ and the limited solid solution on its base, Mo$_{\text{8-x}}$V$_{\text{x}}$Ga$_{41}$ with $x_{\text{max}}=1.9(2)$, have been grown from the high-temperature Ga flux. Magnetization, specific heat and transport measurements confirm bulk superconductivity of Mo$_{8}$Ga$_{41}$ with $T_{\text{c}}=9.8$\,K. Analysis of the specific heat data gives evidence of various non-BCS-type features: (i) the normalized specific heat jump at $T_{\text{c}}$, $\Delta{}c_{\text{p}}/\gamma_{\text{N}}T_{\text{c}}=2.83$, exceeds significantly the weak-coupling BCS limit, for which $\Delta{}c_{\text{p}}/\gamma_{\text{N}}T_{\text{c}}=1.43$. From the specific heat data we estimated the electron-phonon coupling constant $\lambda_{\text{ep}}=0.9$, that is consistent with the strong-coupling regime. (ii) The electronic contribution to the specific heat $c_{\text{eS}}/T$ below $T_{\text{c}}$ significantly deviates from the conventional BCS-type behavior, and a power law with $c_{\text{eS}}(T)\propto{}T^3$ is observed above 3\,K instead. All these facts evidence the strong-coupling non-BCS-type behavior of Mo$_{8}$Ga$_{41}$ and the persistence of this regime upon vanadium doping. Future studies of this material are desirable.

\begin{acknowledgements}
The authors acknowledge insightful discussions with Enno Joon and Walter Schnelle, and thank Victor Tafeenko for the help with single-crystal XRD experiments. The work has been supported by the Russian Foundation for Basic Research, grant \#14-03-31181-mol\_a, and by the Mobilitas program of the ESF, grant MTT77. A.A.T. is also grateful for the financial support by the Federal Ministry for Education and Research under the Sofja Kovalevskaya Award of the Alexander von Humboldt Foundation. We acknowledge the use of a STOE STADIVARI single-crystal X-ray diffractometer purchased under the Lomonosov MSU program of development.
\end{acknowledgements}

\bibliography{fulltext}

\begin{thebibliography}{30}%
\makeatletter
\providecommand \@ifxundefined [1]{%
 \@ifx{#1\undefined}
}%
\providecommand \@ifnum [1]{%
 \ifnum #1\expandafter \@firstoftwo
 \else \expandafter \@secondoftwo
 \fi
}%
\providecommand \@ifx [1]{%
 \ifx #1\expandafter \@firstoftwo
 \else \expandafter \@secondoftwo
 \fi
}%
\providecommand \natexlab [1]{#1}%
\providecommand \enquote  [1]{``#1''}%
\providecommand \bibnamefont  [1]{#1}%
\providecommand \bibfnamefont [1]{#1}%
\providecommand \citenamefont [1]{#1}%
\providecommand \href@noop [0]{\@secondoftwo}%
\providecommand \href [0]{\begingroup \@sanitize@url \@href}%
\providecommand \@href[1]{\@@startlink{#1}\@@href}%
\providecommand \@@href[1]{\endgroup#1\@@endlink}%
\providecommand \@sanitize@url [0]{\catcode `\\12\catcode `\$12\catcode
  `\&12\catcode `\#12\catcode `\^12\catcode `\_12\catcode `\%12\relax}%
\providecommand \@@startlink[1]{}%
\providecommand \@@endlink[0]{}%
\providecommand \url  [0]{\begingroup\@sanitize@url \@url }%
\providecommand \@url [1]{\endgroup\@href {#1}{\urlprefix }}%
\providecommand \urlprefix  [0]{URL }%
\providecommand \Eprint [0]{\href }%
\providecommand \doibase [0]{http://dx.doi.org/}%
\providecommand \selectlanguage [0]{\@gobble}%
\providecommand \bibinfo  [0]{\@secondoftwo}%
\providecommand \bibfield  [0]{\@secondoftwo}%
\providecommand \translation [1]{[#1]}%
\providecommand \BibitemOpen [0]{}%
\providecommand \bibitemStop [0]{}%
\providecommand \bibitemNoStop [0]{.\EOS\space}%
\providecommand \EOS [0]{\spacefactor3000\relax}%
\providecommand \BibitemShut  [1]{\csname bibitem#1\endcsname}%
\let\auto@bib@innerbib\@empty
\bibitem [{\citenamefont {Poole}(2000)}]{HSC}%
  \BibitemOpen
  \bibfield  {author} {\bibinfo {author} {\bibfnamefont {C.~P.}\ \bibnamefont
  {Poole}},\ }\href@noop {} {\emph {\bibinfo {title} {Handbook of
  superconductivity}}}\ (\bibinfo  {publisher} {Academic Press},\ \bibinfo
  {address} {London},\ \bibinfo {year} {2000})\BibitemShut {NoStop}%
\bibitem [{\citenamefont {Morton}\ \emph {et~al.}(1971)\citenamefont {Morton},
  \citenamefont {James}, \citenamefont {Wostenholm}, \citenamefont {Pomfret},
  \citenamefont {Davies},\ and\ \citenamefont {Dykins}}]{MC}%
  \BibitemOpen
  \bibfield  {author} {\bibinfo {author} {\bibfnamefont {N.}~\bibnamefont
  {Morton}}, \bibinfo {author} {\bibfnamefont {B.~W.}\ \bibnamefont {James}},
  \bibinfo {author} {\bibfnamefont {G.~H.}\ \bibnamefont {Wostenholm}},
  \bibinfo {author} {\bibfnamefont {D.~G.}\ \bibnamefont {Pomfret}}, \bibinfo
  {author} {\bibfnamefont {M.~R.}\ \bibnamefont {Davies}}, \ and\ \bibinfo
  {author} {\bibfnamefont {J.~L.}\ \bibnamefont {Dykins}},\ }\href@noop {}
  {\bibfield  {journal} {\bibinfo  {journal} {J. Less-Common Met.}\ }\textbf
  {\bibinfo {volume} {25}},\ \bibinfo {pages} {97} (\bibinfo {year}
  {1971})}\BibitemShut {NoStop}%
\bibitem [{\citenamefont {Bauer}\ \emph {et~al.}(2010)\citenamefont {Bauer},
  \citenamefont {Rogl}, \citenamefont {Chen}, \citenamefont {Khan},
  \citenamefont {Michor}, \citenamefont {Hilscher}, \citenamefont {Royanian},
  \citenamefont {Kumagai}, \citenamefont {Li}, \citenamefont {Li},
  \citenamefont {Podloucky},\ and\ \citenamefont {Rogl}}]{b1}%
  \BibitemOpen
  \bibfield  {author} {\bibinfo {author} {\bibfnamefont {E.}~\bibnamefont
  {Bauer}}, \bibinfo {author} {\bibfnamefont {G.}~\bibnamefont {Rogl}},
  \bibinfo {author} {\bibfnamefont {X.~Q.}\ \bibnamefont {Chen}}, \bibinfo
  {author} {\bibfnamefont {R.~T.}\ \bibnamefont {Khan}}, \bibinfo {author}
  {\bibfnamefont {H.}~\bibnamefont {Michor}}, \bibinfo {author} {\bibfnamefont
  {G.}~\bibnamefont {Hilscher}}, \bibinfo {author} {\bibfnamefont
  {E.}~\bibnamefont {Royanian}}, \bibinfo {author} {\bibfnamefont
  {K.}~\bibnamefont {Kumagai}}, \bibinfo {author} {\bibfnamefont {D.~Z.}\
  \bibnamefont {Li}}, \bibinfo {author} {\bibfnamefont {Y.~Y.}\ \bibnamefont
  {Li}}, \bibinfo {author} {\bibfnamefont {R.}~\bibnamefont {Podloucky}}, \
  and\ \bibinfo {author} {\bibfnamefont {P.}~\bibnamefont {Rogl}},\ }\href@noop
  {} {\bibfield  {journal} {\bibinfo  {journal} {Phys. Rev. B}\ }\textbf
  {\bibinfo {volume} {82}},\ \bibinfo {pages} {064511} (\bibinfo {year}
  {2010})}\BibitemShut {NoStop}%
\bibitem [{\citenamefont {Karki}\ \emph {et~al.}(2010)\citenamefont {Karki},
  \citenamefont {Xiong}, \citenamefont {Vekhter}, \citenamefont {Browne},
  \citenamefont {Adams}, \citenamefont {Young}, \citenamefont {Thomas},
  \citenamefont {Chan}, \citenamefont {Kim},\ and\ \citenamefont
  {Prozorov}}]{k1}%
  \BibitemOpen
  \bibfield  {author} {\bibinfo {author} {\bibfnamefont {A.~B.}\ \bibnamefont
  {Karki}}, \bibinfo {author} {\bibfnamefont {Y.~M.}\ \bibnamefont {Xiong}},
  \bibinfo {author} {\bibfnamefont {I.}~\bibnamefont {Vekhter}}, \bibinfo
  {author} {\bibfnamefont {D.}~\bibnamefont {Browne}}, \bibinfo {author}
  {\bibfnamefont {P.~W.}\ \bibnamefont {Adams}}, \bibinfo {author}
  {\bibfnamefont {D.~P.}\ \bibnamefont {Young}}, \bibinfo {author}
  {\bibfnamefont {K.~R.}\ \bibnamefont {Thomas}}, \bibinfo {author}
  {\bibfnamefont {J.~Y.}\ \bibnamefont {Chan}}, \bibinfo {author}
  {\bibfnamefont {H.}~\bibnamefont {Kim}}, \ and\ \bibinfo {author}
  {\bibfnamefont {R.}~\bibnamefont {Prozorov}},\ }\href@noop {} {\bibfield
  {journal} {\bibinfo  {journal} {Phys. Rev. B}\ }\textbf {\bibinfo {volume}
  {82}},\ \bibinfo {pages} {064512} (\bibinfo {year} {2010})}\BibitemShut
  {NoStop}%
\bibitem [{\citenamefont {Bonalde}\ \emph {et~al.}(2011)\citenamefont
  {Bonalde}, \citenamefont {Kim}, \citenamefont {Prozorov}, \citenamefont
  {Rojas}, \citenamefont {Rogl},\ and\ \citenamefont {Bauer}}]{b3}%
  \BibitemOpen
  \bibfield  {author} {\bibinfo {author} {\bibfnamefont {I.}~\bibnamefont
  {Bonalde}}, \bibinfo {author} {\bibfnamefont {H.}~\bibnamefont {Kim}},
  \bibinfo {author} {\bibfnamefont {R.}~\bibnamefont {Prozorov}}, \bibinfo
  {author} {\bibfnamefont {C.}~\bibnamefont {Rojas}}, \bibinfo {author}
  {\bibfnamefont {P.}~\bibnamefont {Rogl}}, \ and\ \bibinfo {author}
  {\bibfnamefont {E.}~\bibnamefont {Bauer}},\ }\href@noop {} {\bibfield
  {journal} {\bibinfo  {journal} {Phys. Rev. B}\ }\textbf {\bibinfo {volume}
  {84}},\ \bibinfo {pages} {134506} (\bibinfo {year} {2011})}\BibitemShut
  {NoStop}%
\bibitem [{\citenamefont {Bauer}\ \emph {et~al.}(2014)\citenamefont {Bauer},
  \citenamefont {Sekine}, \citenamefont {Sai}, \citenamefont {Rogl},
  \citenamefont {Biswas},\ and\ \citenamefont {Amato}}]{b2}%
  \BibitemOpen
  \bibfield  {author} {\bibinfo {author} {\bibfnamefont {E.}~\bibnamefont
  {Bauer}}, \bibinfo {author} {\bibfnamefont {C.}~\bibnamefont {Sekine}},
  \bibinfo {author} {\bibfnamefont {U.}~\bibnamefont {Sai}}, \bibinfo {author}
  {\bibfnamefont {P.}~\bibnamefont {Rogl}}, \bibinfo {author} {\bibfnamefont
  {P.~K.}\ \bibnamefont {Biswas}}, \ and\ \bibinfo {author} {\bibfnamefont
  {A.}~\bibnamefont {Amato}},\ }\href@noop {} {\bibfield  {journal} {\bibinfo
  {journal} {Phys. Rev. B}\ }\textbf {\bibinfo {volume} {90}},\ \bibinfo
  {pages} {054522} (\bibinfo {year} {2014})}\BibitemShut {NoStop}%
\bibitem [{\citenamefont {Candolfi}\ \emph {et~al.}(2007)\citenamefont
  {Candolfi}, \citenamefont {Lenoir}, \citenamefont {Dauscher}, \citenamefont
  {Bellouard}, \citenamefont {Hejtm{\'a}nek}, \citenamefont
  {{\v{S}}antav{\'a}},\ and\ \citenamefont {Tobola}}]{m1}%
  \BibitemOpen
  \bibfield  {author} {\bibinfo {author} {\bibfnamefont {C.}~\bibnamefont
  {Candolfi}}, \bibinfo {author} {\bibfnamefont {B.}~\bibnamefont {Lenoir}},
  \bibinfo {author} {\bibfnamefont {A.}~\bibnamefont {Dauscher}}, \bibinfo
  {author} {\bibfnamefont {C.}~\bibnamefont {Bellouard}}, \bibinfo {author}
  {\bibfnamefont {J.}~\bibnamefont {Hejtm{\'a}nek}}, \bibinfo {author}
  {\bibfnamefont {E.}~\bibnamefont {{\v{S}}antav{\'a}}}, \ and\ \bibinfo
  {author} {\bibfnamefont {J.}~\bibnamefont {Tobola}},\ }\href@noop {}
  {\bibfield  {journal} {\bibinfo  {journal} {Phys. Rev. Lett.}\ }\textbf
  {\bibinfo {volume} {99}},\ \bibinfo {pages} {037006} (\bibinfo {year}
  {2007})}\BibitemShut {NoStop}%
\bibitem [{\citenamefont {Tran}\ \emph
  {et~al.}(2008{\natexlab{a}})\citenamefont {Tran}, \citenamefont {Miiller},\
  and\ \citenamefont {Bukowski}}]{m2}%
  \BibitemOpen
  \bibfield  {author} {\bibinfo {author} {\bibfnamefont {V.~H.}\ \bibnamefont
  {Tran}}, \bibinfo {author} {\bibfnamefont {W.}~\bibnamefont {Miiller}}, \
  and\ \bibinfo {author} {\bibfnamefont {Z.}~\bibnamefont {Bukowski}},\
  }\href@noop {} {\bibfield  {journal} {\bibinfo  {journal} {Phys. Rev. Lett.}\
  }\textbf {\bibinfo {volume} {100}},\ \bibinfo {pages} {137004} (\bibinfo
  {year} {2008}{\natexlab{a}})}\BibitemShut {NoStop}%
\bibitem [{\citenamefont {Khasanov}\ \emph {et~al.}(2008)\citenamefont
  {Khasanov}, \citenamefont {Klamut}, \citenamefont {Shengelaya}, \citenamefont
  {Bukowski}, \citenamefont {Savi{\'c}}, \citenamefont {Baines},\ and\
  \citenamefont {Keller}}]{m4}%
  \BibitemOpen
  \bibfield  {author} {\bibinfo {author} {\bibfnamefont {R.}~\bibnamefont
  {Khasanov}}, \bibinfo {author} {\bibfnamefont {P.~W.}\ \bibnamefont
  {Klamut}}, \bibinfo {author} {\bibfnamefont {A.}~\bibnamefont {Shengelaya}},
  \bibinfo {author} {\bibfnamefont {Z.}~\bibnamefont {Bukowski}}, \bibinfo
  {author} {\bibfnamefont {I.~M.}\ \bibnamefont {Savi{\'c}}}, \bibinfo {author}
  {\bibfnamefont {C.}~\bibnamefont {Baines}}, \ and\ \bibinfo {author}
  {\bibfnamefont {H.}~\bibnamefont {Keller}},\ }\href@noop {} {\bibfield
  {journal} {\bibinfo  {journal} {Phys. Rev. B}\ }\textbf {\bibinfo {volume}
  {78}},\ \bibinfo {pages} {014502} (\bibinfo {year} {2008})}\BibitemShut
  {NoStop}%
\bibitem [{\citenamefont {Khasanov}\ \emph {et~al.}(2010)\citenamefont
  {Khasanov}, \citenamefont {Shengelaya}, \citenamefont {Savi{\'c}},
  \citenamefont {Baines},\ and\ \citenamefont {Keller}}]{m5}%
  \BibitemOpen
  \bibfield  {author} {\bibinfo {author} {\bibfnamefont {R.}~\bibnamefont
  {Khasanov}}, \bibinfo {author} {\bibfnamefont {A.}~\bibnamefont
  {Shengelaya}}, \bibinfo {author} {\bibfnamefont {I.~M.}\ \bibnamefont
  {Savi{\'c}}}, \bibinfo {author} {\bibfnamefont {C.}~\bibnamefont {Baines}}, \
  and\ \bibinfo {author} {\bibfnamefont {H.}~\bibnamefont {Keller}},\
  }\href@noop {} {\bibfield  {journal} {\bibinfo  {journal} {Phys. Rev. B}\
  }\textbf {\bibinfo {volume} {82}},\ \bibinfo {pages} {016501} (\bibinfo
  {year} {2010})}\BibitemShut {NoStop}%
\bibitem [{\citenamefont {Tran}\ \emph
  {et~al.}(2008{\natexlab{b}})\citenamefont {Tran}, \citenamefont {Hillier},
  \citenamefont {Adroja},\ and\ \citenamefont {Bukowski}}]{m6}%
  \BibitemOpen
  \bibfield  {author} {\bibinfo {author} {\bibfnamefont {V.~H.}\ \bibnamefont
  {Tran}}, \bibinfo {author} {\bibfnamefont {A.~D.}\ \bibnamefont {Hillier}},
  \bibinfo {author} {\bibfnamefont {D.~T.}\ \bibnamefont {Adroja}}, \ and\
  \bibinfo {author} {\bibfnamefont {Z.}~\bibnamefont {Bukowski}},\ }\href@noop
  {} {\bibfield  {journal} {\bibinfo  {journal} {Phys. Rev. B}\ }\textbf
  {\bibinfo {volume} {78}},\ \bibinfo {pages} {172505} (\bibinfo {year}
  {2008}{\natexlab{b}})}\BibitemShut {NoStop}%
\bibitem [{\citenamefont {Tran}\ \emph {et~al.}(2010)\citenamefont {Tran},
  \citenamefont {Hillier},\ and\ \citenamefont {Adroja}}]{m7}%
  \BibitemOpen
  \bibfield  {author} {\bibinfo {author} {\bibfnamefont {V.~H.}\ \bibnamefont
  {Tran}}, \bibinfo {author} {\bibfnamefont {A.~D.}\ \bibnamefont {Hillier}}, \
  and\ \bibinfo {author} {\bibfnamefont {D.~T.}\ \bibnamefont {Adroja}},\
  }\href@noop {} {\bibfield  {journal} {\bibinfo  {journal} {Phys. Rev. B}\
  }\textbf {\bibinfo {volume} {82}},\ \bibinfo {pages} {016502} (\bibinfo
  {year} {2010})}\BibitemShut {NoStop}%
\bibitem [{\citenamefont {Dong}\ \emph {et~al.}(2014)\citenamefont {Dong},
  \citenamefont {Pan}, \citenamefont {Zhang}, \citenamefont {Hong},
  \citenamefont {He}, \citenamefont {Zhou}, \citenamefont {Dong},\ and\
  \citenamefont {Li}}]{m8}%
  \BibitemOpen
  \bibfield  {author} {\bibinfo {author} {\bibfnamefont {W.~N.}\ \bibnamefont
  {Dong}}, \bibinfo {author} {\bibfnamefont {J.}~\bibnamefont {Pan}}, \bibinfo
  {author} {\bibfnamefont {J.}~\bibnamefont {Zhang}}, \bibinfo {author}
  {\bibfnamefont {X.~C.}\ \bibnamefont {Hong}}, \bibinfo {author}
  {\bibfnamefont {L.~P.}\ \bibnamefont {He}}, \bibinfo {author} {\bibfnamefont
  {S.~Y.}\ \bibnamefont {Zhou}}, \bibinfo {author} {\bibfnamefont {J.~K.}\
  \bibnamefont {Dong}}, \ and\ \bibinfo {author} {\bibfnamefont {S.~Y.}\
  \bibnamefont {Li}},\ }\href@noop {} {\bibfield  {journal} {\bibinfo
  {journal} {Solid State Communications}\ }\textbf {\bibinfo {volume} {195}},\
  \bibinfo {pages} {84} (\bibinfo {year} {2014})}\BibitemShut {NoStop}%
\bibitem [{\citenamefont {Bezinge}\ and\ \citenamefont {Yvon}(1984)}]{m3}%
  \BibitemOpen
  \bibfield  {author} {\bibinfo {author} {\bibfnamefont {A.}~\bibnamefont
  {Bezinge}}\ and\ \bibinfo {author} {\bibfnamefont {K.}~\bibnamefont {Yvon}},\
  }\href@noop {} {\bibfield  {journal} {\bibinfo  {journal} {J. Less-Common
  Met.}\ }\textbf {\bibinfo {volume} {99}},\ \bibinfo {pages} {L27} (\bibinfo
  {year} {1984})}\BibitemShut {NoStop}%
\bibitem [{\citenamefont {Yvon}(1975)}]{m0}%
  \BibitemOpen
  \bibfield  {author} {\bibinfo {author} {\bibfnamefont {K.}~\bibnamefont
  {Yvon}},\ }\href@noop {} {\bibfield  {journal} {\bibinfo  {journal} {Acta
  Cryst.}\ }\textbf {\bibinfo {volume} {B31}},\ \bibinfo {pages} {117}
  (\bibinfo {year} {1975})}\BibitemShut {NoStop}%
\bibitem [{\citenamefont {Pet{\v{r}}{\'i}{\v{c}}ek}\ \emph
  {et~al.}(2014)\citenamefont {Pet{\v{r}}{\'i}{\v{c}}ek}, \citenamefont
  {Du{\v{s}}ek},\ and\ \citenamefont {Palatinus}}]{jana}%
  \BibitemOpen
  \bibfield  {author} {\bibinfo {author} {\bibfnamefont {V.}~\bibnamefont
  {Pet{\v{r}}{\'i}{\v{c}}ek}}, \bibinfo {author} {\bibfnamefont
  {M.}~\bibnamefont {Du{\v{s}}ek}}, \ and\ \bibinfo {author} {\bibfnamefont
  {L.}~\bibnamefont {Palatinus}},\ }\href@noop {} {\bibfield  {journal}
  {\bibinfo  {journal} {Z. Kristallogr.}\ }\textbf {\bibinfo {volume}
  {229(5)}},\ \bibinfo {pages} {345} (\bibinfo {year} {2014})}\BibitemShut
  {NoStop}%
\bibitem [{SM()}]{SM}%
  \BibitemOpen
  \href@noop {} {}\bibinfo {note} {Tables~S1, S2 and S3 are presented as
  Supplemental materials}\BibitemShut {NoStop}%
\bibitem [{\citenamefont {Burla}\ \emph {et~al.}(2003)\citenamefont {Burla},
  \citenamefont {Camalli}, \citenamefont {Carrozzini}, \citenamefont
  {Cascarano}, \citenamefont {Giacovazzo}, \citenamefont {Polidori},\ and\
  \citenamefont {Spagna}}]{sir}%
  \BibitemOpen
  \bibfield  {author} {\bibinfo {author} {\bibfnamefont {M.~C.}\ \bibnamefont
  {Burla}}, \bibinfo {author} {\bibfnamefont {M.}~\bibnamefont {Camalli}},
  \bibinfo {author} {\bibfnamefont {B.}~\bibnamefont {Carrozzini}}, \bibinfo
  {author} {\bibfnamefont {G.~L.}\ \bibnamefont {Cascarano}}, \bibinfo {author}
  {\bibfnamefont {C.}~\bibnamefont {Giacovazzo}}, \bibinfo {author}
  {\bibfnamefont {G.}~\bibnamefont {Polidori}}, \ and\ \bibinfo {author}
  {\bibfnamefont {R.}~\bibnamefont {Spagna}},\ }\href@noop {} {\bibfield
  {journal} {\bibinfo  {journal} {J. Appl. Cryst.}\ }\textbf {\bibinfo {volume}
  {36}},\ \bibinfo {pages} {1103} (\bibinfo {year} {2003})}\BibitemShut
  {NoStop}%
\bibitem [{\citenamefont {Gelato}\ and\ \citenamefont
  {Parth{\'e}}(1987)}]{tidy}%
  \BibitemOpen
  \bibfield  {author} {\bibinfo {author} {\bibfnamefont {L.~M.}\ \bibnamefont
  {Gelato}}\ and\ \bibinfo {author} {\bibfnamefont {E.}~\bibnamefont
  {Parth{\'e}}},\ }\href@noop {} {\bibfield  {journal} {\bibinfo  {journal} {J.
  Appl. Cryst.}\ }\textbf {\bibinfo {volume} {20}},\ \bibinfo {pages} {139}
  (\bibinfo {year} {1987})}\BibitemShut {NoStop}%
\bibitem [{\citenamefont {Perdew}\ and\ \citenamefont {Wang}(1992)}]{lda}%
  \BibitemOpen
  \bibfield  {author} {\bibinfo {author} {\bibfnamefont {J.~P.}\ \bibnamefont
  {Perdew}}\ and\ \bibinfo {author} {\bibfnamefont {Y.}~\bibnamefont {Wang}},\
  }\href@noop {} {\bibfield  {journal} {\bibinfo  {journal} {Phys. Rev. B}\
  }\textbf {\bibinfo {volume} {45}},\ \bibinfo {pages} {13244} (\bibinfo {year}
  {1992})}\BibitemShut {NoStop}%
\bibitem [{\citenamefont {Koepernik}\ and\ \citenamefont
  {Eschrig}(1999)}]{fplo}%
  \BibitemOpen
  \bibfield  {author} {\bibinfo {author} {\bibfnamefont {K.}~\bibnamefont
  {Koepernik}}\ and\ \bibinfo {author} {\bibfnamefont {H.}~\bibnamefont
  {Eschrig}},\ }\href@noop {} {\bibfield  {journal} {\bibinfo  {journal} {Phys.
  Rev. B}\ }\textbf {\bibinfo {volume} {59}},\ \bibinfo {pages} {1743}
  (\bibinfo {year} {1999})}\BibitemShut {NoStop}%
\bibitem [{\citenamefont {Bl$\ddot{\text{o}}$chl}\ \emph
  {et~al.}(1994)\citenamefont {Bl$\ddot{\text{o}}$chl}, \citenamefont
  {Jepsen},\ and\ \citenamefont {Andersen}}]{tetr}%
  \BibitemOpen
  \bibfield  {author} {\bibinfo {author} {\bibfnamefont {P.~E.}\ \bibnamefont
  {Bl$\ddot{\text{o}}$chl}}, \bibinfo {author} {\bibfnamefont {O.}~\bibnamefont
  {Jepsen}}, \ and\ \bibinfo {author} {\bibfnamefont {O.~K.}\ \bibnamefont
  {Andersen}},\ }\href@noop {} {\bibfield  {journal} {\bibinfo  {journal}
  {Phys. Rev. B}\ }\textbf {\bibinfo {volume} {49}},\ \bibinfo {pages} {16223}
  (\bibinfo {year} {1994})}\BibitemShut {NoStop}%
\bibitem [{\citenamefont {Gunnarsson}\ \emph {et~al.}(2003)\citenamefont
  {Gunnarsson}, \citenamefont {Calandra},\ and\ \citenamefont {Han}}]{r1}%
  \BibitemOpen
  \bibfield  {author} {\bibinfo {author} {\bibfnamefont {O.}~\bibnamefont
  {Gunnarsson}}, \bibinfo {author} {\bibfnamefont {M.}~\bibnamefont
  {Calandra}}, \ and\ \bibinfo {author} {\bibfnamefont {J.~E.}\ \bibnamefont
  {Han}},\ }\href@noop {} {\bibfield  {journal} {\bibinfo  {journal} {Rev. Mod.
  Phys.}\ }\textbf {\bibinfo {volume} {75}},\ \bibinfo {pages} {4} (\bibinfo
  {year} {2003})}\BibitemShut {NoStop}%
\bibitem [{\citenamefont {Fisk}\ and\ \citenamefont {Webb}(1976)}]{r2}%
  \BibitemOpen
  \bibfield  {author} {\bibinfo {author} {\bibfnamefont {Z.}~\bibnamefont
  {Fisk}}\ and\ \bibinfo {author} {\bibfnamefont {G.~W.}\ \bibnamefont
  {Webb}},\ }\href@noop {} {\bibfield  {journal} {\bibinfo  {journal} {Phys.
  Rev. Lett.}\ }\textbf {\bibinfo {volume} {36}},\ \bibinfo {pages} {1084}
  (\bibinfo {year} {1976})}\BibitemShut {NoStop}%
\bibitem [{\citenamefont {Bain}\ and\ \citenamefont {Berry}(2008)}]{dia}%
  \BibitemOpen
  \bibfield  {author} {\bibinfo {author} {\bibfnamefont {G.~A.}\ \bibnamefont
  {Bain}}\ and\ \bibinfo {author} {\bibfnamefont {J.~F.}\ \bibnamefont
  {Berry}},\ }\href@noop {} {\bibfield  {journal} {\bibinfo  {journal} {J.
  Chem. Educ.}\ }\textbf {\bibinfo {volume} {85(4)}},\ \bibinfo {pages} {532}
  (\bibinfo {year} {2008})}\BibitemShut {NoStop}%
\bibitem [{\citenamefont {McMillan}(1968)}]{MM}%
  \BibitemOpen
  \bibfield  {author} {\bibinfo {author} {\bibfnamefont {W.~L.}\ \bibnamefont
  {McMillan}},\ }\href@noop {} {\bibfield  {journal} {\bibinfo  {journal}
  {Phys. Rev.}\ }\textbf {\bibinfo {volume} {167}},\ \bibinfo {pages} {331}
  (\bibinfo {year} {1968})}\BibitemShut {NoStop}%
\bibitem [{\citenamefont {Saint-James}\ \emph {et~al.}(1969)\citenamefont
  {Saint-James}, \citenamefont {Sarma},\ and\ \citenamefont {Thomas}}]{hc3}%
  \BibitemOpen
  \bibfield  {author} {\bibinfo {author} {\bibfnamefont {D.}~\bibnamefont
  {Saint-James}}, \bibinfo {author} {\bibfnamefont {G.}~\bibnamefont {Sarma}},
  \ and\ \bibinfo {author} {\bibfnamefont {E.~J.}\ \bibnamefont {Thomas}},\
  }\href@noop {} {\emph {\bibinfo {title} {Type II superconductivity}}}\
  (\bibinfo  {publisher} {Elsevier Science \& Technology},\ \bibinfo {year}
  {1969})\BibitemShut {NoStop}%
\bibitem [{\citenamefont {Hempstead}\ and\ \citenamefont {Kim}(1964)}]{hc32}%
  \BibitemOpen
  \bibfield  {author} {\bibinfo {author} {\bibfnamefont {C.~F.}\ \bibnamefont
  {Hempstead}}\ and\ \bibinfo {author} {\bibfnamefont {Y.~B.}\ \bibnamefont
  {Kim}},\ }\href@noop {} {\bibfield  {journal} {\bibinfo  {journal} {Phys.
  Rev. Lett.}\ }\textbf {\bibinfo {volume} {12}},\ \bibinfo {pages} {145}
  (\bibinfo {year} {1964})}\BibitemShut {NoStop}%
\bibitem [{\citenamefont {Werthamer}\ \emph {et~al.}(1966)\citenamefont
  {Werthamer}, \citenamefont {Helfand},\ and\ \citenamefont {Honenberg}}]{whh}%
  \BibitemOpen
  \bibfield  {author} {\bibinfo {author} {\bibfnamefont {N.~R.}\ \bibnamefont
  {Werthamer}}, \bibinfo {author} {\bibfnamefont {E.}~\bibnamefont {Helfand}},
  \ and\ \bibinfo {author} {\bibfnamefont {P.~C.}\ \bibnamefont {Honenberg}},\
  }\href@noop {} {\bibfield  {journal} {\bibinfo  {journal} {Phys. Rev.}\
  }\textbf {\bibinfo {volume} {147}},\ \bibinfo {pages} {295} (\bibinfo {year}
  {1966})}\BibitemShut {NoStop}%
\bibitem [{\citenamefont {Viklund}\ \emph {et~al.}(2001)\citenamefont
  {Viklund}, \citenamefont {Svensson}, \citenamefont {Hull}, \citenamefont
  {Simak}, \citenamefont {Berastegui},\ and\ \citenamefont
  {H$\ddot{\text{a}}$ussermann}}]{h1}%
  \BibitemOpen
  \bibfield  {author} {\bibinfo {author} {\bibfnamefont {P.}~\bibnamefont
  {Viklund}}, \bibinfo {author} {\bibfnamefont {C.}~\bibnamefont {Svensson}},
  \bibinfo {author} {\bibfnamefont {S.}~\bibnamefont {Hull}}, \bibinfo {author}
  {\bibfnamefont {S.~I.}\ \bibnamefont {Simak}}, \bibinfo {author}
  {\bibfnamefont {P.}~\bibnamefont {Berastegui}}, \ and\ \bibinfo {author}
  {\bibfnamefont {U.}~\bibnamefont {H$\ddot{\text{a}}$ussermann}},\ }\href@noop
  {} {\bibfield  {journal} {\bibinfo  {journal} {Chem. Eur. J.}\ }\textbf
  {\bibinfo {volume} {7(23)}},\ \bibinfo {pages} {5143} (\bibinfo {year}
  {2001})}\BibitemShut {NoStop}%
\end{thebibliography}%
\end{document}